\documentclass[aps,prb,groupedaddress,twocolumn,showpacs]{revtex4}
\usepackage{graphicx}

\begin{document}

\title{ Zero-temperature resistive transition in Josephson-junction arrays
at irrational frustration}

\author{Enzo Granato}

\address{Laborat\'orio Associado de Sensores e Materiais, \\
Instituto Nacional de Pesquisas Espaciais, \\
12245-970 S\~ao Jos\'e dos Campos, SP Brazil}

\begin{abstract}
We use a driven Monte Carlo dynamics in the phase representation
to determine the linear resistivity and current-voltage scaling of
a two-dimensional Josephson-junction array at an irrational flux
quantum per plaquette. The results are consistent with a
phase-coherence transition scenario where the critical temperature
vanishes. The linear resistivity is nonzero at any finite
temperatures but nonlinear behavior sets in at a
temperature-dependent crossover current determined by the thermal
critical exponent. From a dynamic scaling analysis we determine
this critical exponent and the thermally activated behavior of the
linear resistivity. The results are in agreement with  earlier
calculations using the resistively shunted-junction model for the
dynamics of the array. The linear resistivity behavior is
consistent with some experimental results on arrays of
superconducting grains but not on wire networks, which we argue
have been obtained in a current regime above the crossover
current.
\end{abstract}


\maketitle

Most theoretical investigations of the vortex-glass phase in
superconductors have considered model systems where there is a
combined effect of quenched disorder and frustration
\cite{fisher}. However, in artificial Josephson-junction arrays,
frustration without disorder can in principle be introduced by
applying an external magnetic field on a perfect periodic array of
weakly coupled superconducting grains \cite{carini,zant,baek} and
similarly on superconducting wire networks \cite{yu,ling}. The
frustration parameter $f$, the number of flux quantum per
plaquette, is given by $f=\phi/\phi_o$, the ratio of the magnetic
flux through a plaquette $\phi$ to the superconducting flux
quantum $\phi_o = hc/2e$. It can be tuned by varying the strength
of the external field. Frustration effects can be viewed as
resulting from a competition between the underlying periodic
pinning potential of the array and the natural periodicity of the
vortex lattice \cite{teitel}. At a rational value of $f$, the
ground state is a commensurate pinned vortex lattice leading to
discrete symmetries in addition to the continuous $U(1)$ symmetry
of the superconducting order parameter. The resistive transition
is only reasonably well understood for simple rational values of
$f$.

At irrational values of $f$, the resistive behavior is much less
understood since the vortex lattice is now incommensurate with the
periodic array. In early Monte Carlo (MC) simulations \cite{halsey}
the ground state was found to consist of a disordered vortex pattern
lacking long range order which could be regarded as a some sort of
vortex-glass state without quenched disorder. Glassy-like behavior
was indeed observed in these simulations suggesting a possible
superconducting (vortex-glass) transition at finite temperatures.
However, some arguments also suggested that the critical temperature
should vanish \cite{teitel,choi}. Simulations of the current-voltage
scaling using the resistively shunted-junction model for the
dynamics of the array found that the behavior was consistent with an
equilibrium resistive transition where the critical temperature
vanishes \cite{eg96}, similar to the resistive transition described
by the the gauge-glass model in two dimensions \cite{fisher,eg98},
but with different values for the correlation-length critical
exponent $\nu$. The linear resistivity is nonzero at any finite
temperatures but nonlinear behavior sets in at a crossover current
with a temperature dependence determined by the exponent $\nu$. This
zero-temperature transition leads to slow relaxation dynamics where
the correlation length diverges as a power law and the relaxation
time diverges exponentially as the temperature vanishes.

Simulations of the relaxation dynamics \cite{kimlee} found a
behavior analogous to relaxation in supercooled liquids with a
characteristic dynamic crossover temperature rather than an
equilibrium transition temperature, which is not inconsistent with
the zero-temperature transition scenario. On the other hand, a
systematic study by MC simulations \cite{teitelf} of a sequence of
rational values of $f$ converging to the irrational frustration,
using the vortex representation, found two phase transitions at
finite temperatures, a vortex-order transition weakly dependent on
$f$ and a vortex pinning transition at much lower temperatures
varying with $f$, which should correspond to the resistive
transition. These results are in qualitative agreement with MC
simulations using the phase representation of the same model
\cite{tang} but different ground states were found.

More recently, MC simulations for the the specific heat and
relaxation dynamics found an intrinsic finite-size effect
\cite{park}. The corresponding scaling analysis suggested a
zero-temperature transition with a critical exponent $\nu$
consistent with the value obtained initially from current-voltage
scaling \cite{eg96}. However, a study of the low-temperature
configurations for frustrations close the irrational value by MC
simulations in the vortex representation \cite{llkim}, find two
phase transitions consistent with earlier work \cite{teitelf}.

On the experimental side, some results on arrays of superconducting
grains at irrational frustration \cite{carini,zant} are consistent
with the scenario of the zero-temperature resistive transition but
on wire networks \cite{ling,yu}, resistivity scaling showed evidence
of a transition at finite temperature. Recently, resistivity scaling
suggesting a finite temperature transition was also observed in
arrays of superconducting grains \cite{baek}.

In view of these conflicting results, it seems useful to further
investigate the current-voltage scaling for the array at irrational
frustration by studying both the nonlinear and linear resistivity
with an improved method \cite{eg04} taking into account the long
relaxation times. In fact, as found recently, current-voltage
scaling turned out to be quite reliable in determining the
phase-coherence transition even for a model with quenched disorder,
such as the three-dimensional XY-spin glass model \cite{eg04,ly}.
The main question is therefore, if the array at irrational
frustration displays an equilibrium phase-coherence transition at a
nonzero critical temperature into a state with vanishing linear
resistivity or its critical temperature vanishes and the linear
resistivity is finite at nonzero temperatures.

In this work, we investigate the resistivity scaling of
Josephson-junction arrays at a irrational frustration
$f=(3-\sqrt{5})/2$, a golden irrational, using a driven MC
dynamics in the phase representation introduced recently
\cite{eg04}.  The results are consistent with a phase-coherence
transition scenario where the critical temperature vanishes,
$T_c=0$. The linear resistivity is finite at nonzero temperatures
but nonlinear behavior sets in at a temperature-dependent
crossover current  determined by the thermal critical exponent
$\nu$. The results agree  with earlier simulations using the
resistively shunted-junction model for the dynamics of the array
\cite{eg96}. However, with the present MC method we are able to
reach much lower temperatures and current densities, improving the
analysis of resistivity scaling and the estimate of the critical
exponent $\nu$.  We also argue that the finite-temperature
transition found in resistivity measurements on wire networks
\cite{ling,yu}have been obtained in a current regime above the
crossover current.

We consider a two-dimensional Josephson-junction square array
described by the Hamiltonian
\begin{equation}
H=-J_o\sum_{<ij>}\cos(\theta_i -\theta_j-A_{ij}) -J
\sum_i(\theta_i-\theta_{i+x})  \label{model}
\end{equation}
The first term gives the Josephson-coupling energy between nearest
neighbor grains where line integral of the vector potential $A_{ij}$
is constrained to $\sum_{ij}A_{ij} = 2 \pi f$ around each plaquette.
The second term represents the effects of an external driving
current density $J$ applied in the $x$ direction. When $J\ne 0$, the
total energy is unbounded and the system is out of equilibrium. The
lower-energy minima occur at phase differences $\theta_i
-\theta_{i+x}$ which increases with time $t$, leading to a net phase
slippage rate proportional to $< d(\theta_i -\theta_{i+x})/dt>$,
corresponding to the voltage $V_{i,i+x}$. We take the frustration
parameter $f$ equals an irrational number, $f=(3-\sqrt{5})/2$,
related to the Golden Ratio $\Phi=(1+\sqrt{5})/2$ as $f=1-1/\Phi$.
In the numerical simulations we use periodic (fluctuating twist)
boundary conditions on lattices of linear sizes $L$ and
corresponding rational approximations $\Phi=F_{n+1}/F_n$, where
$F_n$ are Fibonacci numbers ($13,21,34,55$), with $L=F_n$.

To study the current-voltage scaling, we use a driven MC dynamics
method \cite{eg04}. The time dependence is obtained by identifying
the MC time as the real time $t$ and we set the unit of time
$dt=1$, corresponding to a complete MC pass through the lattice.
Periodic (fluctuating twist) boundary conditions are used
\cite{saslow}. This boundary condition adds new dynamical
variables, $u_\alpha$ ($\alpha =x$ and $y$), corresponding to a
uniform phase twist between nearest-neighbor sites along the
principal axis directions $\hat x$ and $\hat y$. A MC step
consists of an attempt to change the local phase $\theta_i$  and
the phase twists $u_\alpha$ by fixed amounts, using the Metropolis
algorithm. If the change in energy is $\Delta H$, the trial move
is accepted with probability $ min\{1,\exp(-\Delta H/kT)\}$. The
external current density $J$ in Eq. \ref{model} biases these
changes, leading to a net voltage (phase slippage rate) across the
system given by
\begin{equation}
 V = \frac{1}{L}\frac{d}{dt}  \sum_{j=1}^L(\theta_{1,j} - \theta_{L+1,j}-u_xL),
\end{equation}
in arbitrary units.  The main advantage of this MC method compared
with the Langevin dynamics used earlier \cite{eg96} is that in
principle much longer time scales can be accessed which allows one
to obtain reliable data at much lower temperatures and current
densities. We have determined the electric field $E=V/L$ and
nonlinear resistivity $\rho = E/J$ as a function of the driving
current density $J$ for different temperatures $T$ and different
system sizes. We used typically $ 2 \times 10^5$ MC steps to reach
the nonequilibrium steady state at finite current and equal time
steps to perform time averages with and  additional average over
$4-6$ independent runs.

We have also determined the linear resistivity,
$\rho_L=\lim_{J->0} E/J$, from equilibrium MC simulations. As any
transport coefficient, this quantity can be obtained from
equilibrium fluctuations and therefore can be calculated  in
absence of an imposing driving current ($J=0$). From Kubo formula,
the linear resistivity (resistance in two dimensions) is given in
terms of the equilibrium voltage autocorrelation as
\begin{equation}
\rho_L=\frac{1}{2T} \int d t \langle V(t) V(0) \rangle .
\label{kubo}
\end{equation}
Since the total voltage $V$ is related to the phase difference
across the system $\Delta \theta(t)$ by $V= d\Delta \theta(t)/dt$,
we find more convenient to determine $\rho_L$ from the long-time
equilibrium fluctuations \cite{eg98} of $\Delta \theta(t)$ as
\begin{equation}
\rho_L=\frac{1}{2Tt} \langle (\Delta \theta(t)-\Delta
\theta(0))^2\rangle  , \label{rho}
\end{equation}
which is valid for sufficiently long times $t$. To insure that
only equilibrium fluctuations are considered, the calculations
were performed in two steps. First, simulations using the exchange
MC method (parallel tempering) \cite{nemoto} were used to obtain
equilibrium configurations of the systems at different
temperatures \cite{egunp}. This method is known to reduce
significantly the critical slowing down near the transition
allowing fully equilibration in finite small system sizes. These
configurations were then used as initial states for the driven MC
dynamics process described above, with $J=0$, in order to obtain
the $\rho_L$. The initial states are similar to the
low-temperature states obtained previously \cite{teitelf,llkim}
including thermal excitations. In the parallel-tempering method
\cite{nemoto}, many replicas of the system with different
temperatures are simulated simultaneously and the corresponding
configurations are allowed to be exchanged with a probability
satisfying detailed balance. The equilibration time can be
measured as the average number of MC steps required for each
replica to travel over the whole temperature range. We used
typically $4 \times 10^6$ (parallel tempering) MC steps for
equilibration which is much larger than the estimated
equilibration time in systems with up to $100$ replicas.
Subsequent MC simulations for the linear resistivity obtained from
Eq. \ref{rho} were performed using  $2 \times 10^3$ time averages
for $2 \times 10^5$ MC steps which is much larger than the
equilibrium relaxation time.

\begin{figure}
\includegraphics[bb= 2cm 3cm  19cm   16cm, width=7.5 cm]{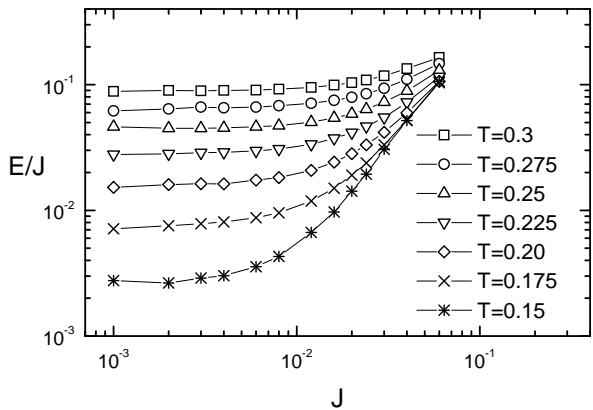}
\caption{ Nonlinear resistivity $E/J$ at different temperatures $T$
for system size $L=55$.}
\end{figure}

Fig. 1a shows the nonlinear resistivity $E/J$  as a function of
temperature for the largest system size.  At small current
densities $J$, the nonlinear resistivity $E/J$ tends to a constant
value, corresponding to the linear resistivity $\rho_L$, which
decreases rapidly with decreasing temperature. For increasing $J$,
the resistivity cross over to a nonlinear behavior at a
characteristic current density $J_{nl}$, which also decreases with
decreasing temperature. To verify that the nonzero values
approached at low currents in Fig. 1 correspond indeed to the
linear resistivity $\rho_L$, we show in Fig. 2 the temperature
dependence of $\rho_L$ obtained without current bias from
Eq.(\ref{rho}) for different system sizes. $\rho_L$ decreases with
system size but approaches nonzero values for the largest system
size. These values are in agreement with the corresponding values
at the lowest current in Fig. 1. Since the behavior of the
$\rho_L$ for the largest system size on the log-linear plot in
Fig. 2 is a straight line, it indicates  an activated Arrhenius
behavior, where the linear resistivity decreases exponentially
with the inverse of temperature with a temperature-independent
energy barrier, estimated as  $E_b \sim 1.07$. Such activated
behavior suggests that the linear resistivity can be very small at
low temperatures but nevertheless remains finite for all
temperatures $T > 0$ and therefore there is no resistive
transition at finite temperatures. However, as will be described
below, the system behaves as if a resistive transition occurs at
zero temperature, corresponding to a phase-coherence transition
where the critical temperature vanishes, $T_c=0$.

The behavior of the linear resistivity can be related to the
equilibrium relaxation time for phase fluctuations. Since the
voltage is the rate of change of the phase, a nonzero $\rho_L$
requires measurements over a time scale $\tau \propto 1/\rho_L$,
corresponding to the relaxation time for phase fluctuations. Thus,
we expect that $\tau$ should also have an activated behavior,
increasing exponentially with the inverse of temperature. To
verify this behavior, we have in addition calculated the
relaxation time $\tau$ for different temperatures from the
autocorrelation function of phase fluctuations $C(t)$ as
\begin{equation}
 \tau=\frac{1}{C(0)^2}\int_0^\infty  dt C(t)
\end{equation}
using MC simulations with $J=0$. The starting configurations were
taken from equilibrium configurations obtained \cite{egunp} with
the parallel tempering MC method \cite{nemoto}. The results shown
on the log-linear plot in Fig. 3 are indeed consistent with an
activated behavior of $\tau$ with an energy barrier $E_b=1.18$ in
reasonable agreement with the value obtained for the linear
resistivity in Fig. 2.

\begin{figure}
\includegraphics[bb= 1cm 3.5cm  19cm   16cm, width=7.5 cm]{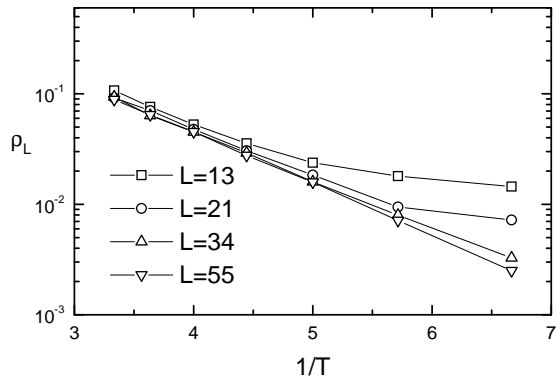}
\caption{ Temperature dependence of the linear resistivity for
different system sizes.}
\end{figure}

The  behavior in Figs. 1, 2 and 3 has the main features associated
with a phase transition that only occurs at zero temperature,
$T_c=0$, similar to the two-dimensional gauge glass model of
disordered superconductors \cite{fisher,eg98}. In this case the
correlation length $\xi$ is finite for $T>0$ but it  increases
with decreasing temperature as $\xi \propto T^{-\nu}$, with $\nu$
a critical exponent.  The divergent correlation length near the
transition determines both the linear an nonlinear resistivity
behavior leading to current-voltage scaling sufficiently close to
the critical temperature and sufficiently small driving current.
To understand in detail the behavior of the linear $\rho_L$ and
nonlinear resistivity $\rho$ we need a scaling theory for the
resistive behavior. If the data satisfy such scaling behavior for
different driving currents and temperatures, the critical
temperature and critical exponents of the underlying equilibrium
transition at $J=0$ can then be determined from the best data
collapse. A detailed scaling theory has been described in the
context of the current-voltage characteristics of vortex-glass
models \cite{fisher} but the arguments should also apply to the
present case. The basic assumption is the existence of a second
order phase transition. Measurable quantities should then scale
with the diverging correlation length $\xi \propto |T-T_c|^{-\nu}$
and relaxation time $\tau$ near the critical point. The nonlinear
resistivity $E/J$ should then satisfy the scaling form
\cite{fisher}
\begin{equation}
T  \frac{E}{J} \tau = g_{\pm}(\frac{J \xi}{T}), \label{scaltc}
\end{equation}
in two-dimensions, where $g_{\pm }(x)$ is a scaling function. The
$+$ and $-$ signs correspond to $T>T_c$ and $T<T_c$, respectively.
If $T_c \ne 0$, then to satisfy such scaling form, the nonlinear
resistivity curves on the log-log plot in Fig. 1 should have a
positive curvature at small $J$, with $E/J$ decreasing with
decreasing $J$ to a temperature dependent value for $T>T_c$ while
for $T<T_c$, the curvature should be negative, with $E/J$
vanishing in the limit $J \rightarrow 0$. The data in Fig. 1 do
not show a change in curvature even for the lowest temperature,
already suggesting the possibility of a resistive transition at
much lower temperatures or at $T_c=0$. However, a full scaling
analysis of the data is required to show that a transition indeed
occur with $T_c=0$. If $T_c=0$, then the correlation length $\xi
\propto T^{- \nu}$ and the linear resistivity $\rho_L$ are both
finite at $T>0$. One can then consider the behavior of the
dimensionless ratio $E/J\rho_L$ which should satisfy the scaling
form
\begin{equation}
\frac{E}{J \rho_{L}}=g(\frac{J}{T^{1+\nu} }) \label{scalt0}
\end{equation}
where $g$ is a scaling function with $g(0) =1$. A crossover from
linear behavior, when $g(x) \sim 1 $, to nonlinear behavior, when
$g(x) >>1$, occurs when $x \sim 1$ which leads to a characteristic
current density at which nonlinear behavior sets in decreasing with
temperatures as a power law, $J_{nl}\propto T/\xi \propto T^{1+\nu
}$. The scaling form in Eq. (\ref{scalt0})contains a single critical
exponent $\nu$ and does not depend on the particular form assumed
for the divergence of the relaxation time $\tau$. However, for
sufficiently low temperatures, the relaxation process is expected to
be thermally activated \cite{fisher} with $\tau \propto
\exp(E_b/kT)$. This corresponds formally to a dynamic exponent $z
\rightarrow \infty$, if power-law behavior is assumed for the
relaxation time $\tau \propto \xi^z$. From the scaling form of
Eq.(\ref{scaltc}), the linear resistivity should scale as $\rho_L
\propto 1/\tau$ and therefore it is also expected to have an
activated behavior, $\tau \propto \exp(-E_b/kT)$.  In general, the
energy barrier $E_b$ also scales with the correlation length as $E_b
\propto \xi^\psi $, which leads to a temperature-dependent barrier
$E_b \propto T^{-\psi \nu }$. A pure Arrhenius behavior corresponds
to $\psi = 0$. The behavior of the nonlinear and linear resistivity
in Figs 1, 2 and the relaxation time in Fig. 3 are quite consistent
with these predictions from the scaling theory of a zero-temperature
transition.

\begin{figure}
\includegraphics[bb= 1cm 3.5cm  19cm   16cm, width=7.5 cm]{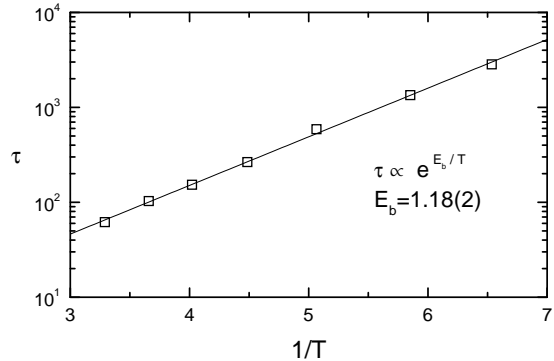}
\caption{ Temperature dependence of the relaxation time $\tau$ of
phase fluctuations for system size $L=55$.}
\end{figure}

If there is a zero-temperature transition, as suggested by the
behaviors in Figs. 1, 2 and 3, then the data for the nonlinear
resistivity should satisfy the scaling form of Eq.(\ref{scalt0}),
if finite-size effects are negligible,  and the best data collapse
provides an estimate of the critical exponent $\nu$. We expect
that finite-size effects are negligible for the largest system
size $L=55$ in Fig. 1 since at this length scale the behavior of
the linear resistivity is roughly independent of the size as can
be seen from Fig. 2. Fig. 4 shows that indeed the data for the
largest system size satisfy this scaling form with $\nu \sim 1.4
\pm 0.2$.

\begin{figure}
\includegraphics[bb= 1cm 3cm  19cm   17cm, width=7.5 cm]{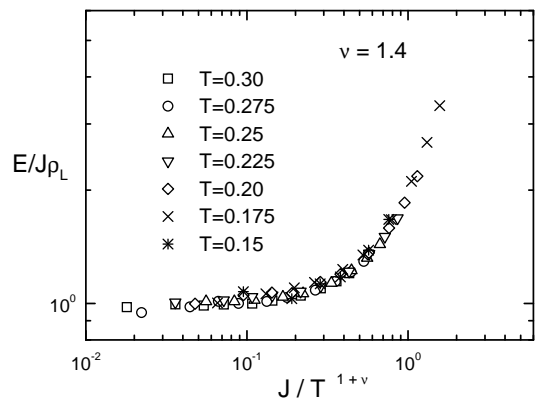}
\caption{ Scaling plot of the nonlinear resistivity in Fig. 1 for
$\nu =1.4$. }
\end{figure}

The nonlinear resistivity should also satisfy the expected
finite-size behavior in smaller system sizes when the correlation
length $\xi$ approaches the system size $L$. According to
finite-size scaling, the scaling function in Eq. (\ref{scalt0}),
should also depend on the dimensionless ratio $L/\xi$ and so to
account for finite-size effects the nonlinear resistivity should
satisfy the scaling form
\begin{equation}
\frac{E}{J \rho_{L}}=\bar{g}(\frac{J}{T^{1+\nu} },L^{1/\nu}T).
\label{scaltL}
\end{equation}
The scaling analysis of the whole nonlinear resistivity data is
rather complicated in this case since the scaling function depends
on two variables. To simplify the analysis \cite{wengel} we first
estimate the temperature and finite-size behavior of the crossover
current density $J_{nl}$ where nonlinear behavior sets in as the
value of $J$ where $E/J \rho_L = C$, a constant. Then, from Eq.
(\ref{scaltL}), the finite-size behavior of $J_{nl}$ can be
expressed in the scaling form
\begin{equation}
J_{nl} L^{(1+\nu)/\nu}=\bar{\bar{g}}(L^{1/\nu}T).
 \label{scaltnl}
\end{equation}
The best data collapse according to the scaling in Eq.
(\ref{scaltnl}) provides an alternative estimate of the critical
exponent $\nu$. Fig. 5 shows that indeed the values of $J_{nl}$ for
different system sizes and temperatures satisfy this scaling form
with $\nu \sim 1.4$, in agreement with the estimate obtained for the
largest system  in Fig. 4 size using Eq. (\ref{scalt0}).

In addition to the standard finite-size effects, which occur when
the correlation length is comparable to the system size, already
taken into account in the scaling form of Eq. (\ref{scaltL}), there
are also intrinsic finite-size effects \cite{park} resulting from
the rational approximations used for the irrational value of $f$.
Since we use rational approximations $\Phi=F_{n+1}/F_n$, where $F_n$
are Fibonacci numbers ($13,21,34,55$), with the system size set to
$L=F_n$, this amounts essentially to have different values of the
frustration, $f_L=1-1/\Phi$, for different system sizes which will
only converge to the correct value $f=(3-\sqrt 5)/2$ in the
infinite-size limit. We have assumed that such effects are
negligible in the above scaling analysis but they should affect our
estimate of the critical exponent $\nu$. In principle, this
intrinsic effect could be taken into account within the
zero-temperature transition scenario by allowing for a
size-dependent critical temperature $T_c(L)$ in the scaling analysis
\cite{park}. Alternatively, we could regard it as a crossover from
the critical behavior at the true irrational frustration
(infinite-size limit) to a phase with an additional small
frustration $\delta f=f_L-f$ which should act as a relevant
perturbation. In this case, the scaling function in Eq.
(\ref{scalt0}) should also depend on the dimensionless ratio
$\xi^2\delta f$ and again a scaling analysis with more than one
variable is required. However, our present numerical data is not
sufficiently accurate to separate this effect from standard
finite-size effects.

\begin{figure}
\includegraphics[bb= 1cm 4cm  19cm   16cm, width=7.5 cm]{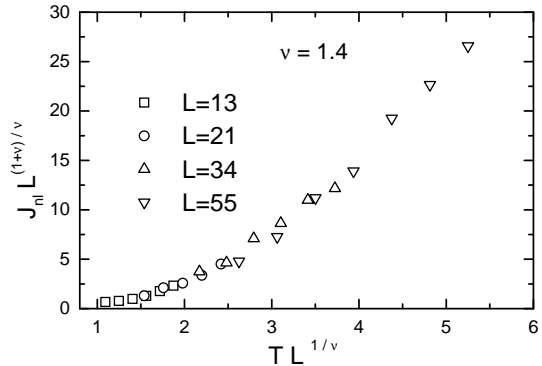}
\caption{ Finite-size scaling plot of the crossover current density
$J_{nl}$ with $\nu =1.4$, for different system sizes $L$.}
\end{figure}

The present results for the linear and nonlinear resistivity of
the array at irrational frustration obtained by the driven MC
dynamics agree with earlier simulations of the current-voltage
scaling using the resistively shunted-junction model for the
dynamics of the array \cite{eg96}, where a zero-temperature
resistive transition was suggested and the critical exponent was
estimated as $\nu =0.9(2)$. Although  the later model is expected
to be a more realistic description for the dynamics of the array,
the value of the static critical exponent $\nu$ should be the same
for both models. In general, the dynamic exponent $z$ may depend
on the particular dynamics but since the relaxation time $\tau$ is
found to diverge exponentially for decreasing temperature it
corresponds to $z \rightarrow \infty$ for both dynamics. The
present estimate of $\nu=1.4(2)$, however, should be more reliable
since it considers much lower temperatures and current densities
and larger system size. Interestingly, similar behavior for the
resistive transition has been found both numerically and
experimentally for two-dimensional disordered superconductors in a
magnetic field described  as a gauge-glass model
\cite{fisher,eg98} but with different value for critical exponent
$\nu \sim 2$. It should be noted however that the actual ground
state at irrational frustration (without disorder) can be quite
different, as the self similar structure which has already been
proposed \cite{yu,kolahchi}. As would be expected, the different
nature of ground state leads to the different values of the
critical exponent $\nu$.

Although the above scaling analysis is consistent with a zero
temperature transition, on pure numerical grounds the data in Figs 1
and 2 can not complete ruled out a vortex-order or a phase-coherence
transition at temperatures much lower than $T = 0.15$. In fact,
phase-coherence transitions were found in MC simulations using the
Coulomb-gas presentation \cite{teitelf} at temperatures as low as $T
\sim 0.03$ for the sequence of rational approximations $f_L$ of the
irrational $f$ but since they show considerable variation with $f_L$
it is not clear if it will remain nonzero in the large size limit.
However, the lowest temperature in Figs 1 and 2  is already much
smaller than the apparent freezing temperature $T_f \sim 0.25$
observed in earlier MC simulations \cite{halsey}. Below $T_f$, a
nonzero Edwards-Anderson order parameter $q(t)=<\vec S_i>^2$, was
observed, where $\vec S = (\cos \theta, \sin \theta ) $ and the
average was taken over the simulation times $t$. Although this could
suggest a diverging relaxation time $\tau \propto \int_o^\infty q(t)
dt $ near a finite temperature $T_c \sim T_f$, such long relaxation
time can also result from a zero-temperature transition ($T_c=0$) as
suggested by the above scaling analysis since in this case $\tau$
diverges exponentially with decreasing temperature, $\tau \propto
\exp(E_b/kT)$, as shown in Fig. 3. For low enough temperatures,
$\tau$ will eventually be larger than any simulation or experimental
measuring time scale and an apparent (time dependent) freezing
transition could occur depending on the particular dynamics and
system size.

Some experimental results on arrays of superconducting grains at
irrational frustration \cite{carini,zant} are consistent with the
scenario of a zero-temperature resistive transition since even at
the lowest temperatures a zero-resistance state was not observed in
these experiments. On the other hand, current-voltage scaling
analysis of experimental data on wire networks \cite{ling,yu} was
found to be consistent with  a resistive transition at finite
temperature. We note, however, that although the equilibrium
behavior of wire networks can be described by the same model of Eq.
\ref{model}, the nonlinear dynamical behavior may be quite different
since the nodes of the network are connected by continuous
superconducting wires, instead of weak links, leading to additional
larger energy barriers for vortex motion, not included in the model,
and consequently larger phase-coherence length $\xi$ and relaxation
time $\tau$ when compared with weak links \cite{zant90}. In this
case, the characteristic crossover current to the linear resistivity
regime at low temperatures due to thermal fluctuations, $J_{nl}
\propto kT/\xi$, expected in the zero-temperature transition
scenario, may only occur at current scales too small to be detected
experimentally. Thus the resistive behavior is observed in a current
regime at higher currents where it follows the mean-field theory
result \cite{parisi} where a vortex-glass transition is possible at
finite temperatures. However, the zero-temperature resistive could
in principle be observed in specially prepared wire networks in the
weak coupling regime where the additional energy barrier for vortex
motion can be minimized \cite{giroud}. Other effects, such as weak
disorder, which is inevitably present in both experimental systems,
should also be considered. It could provide a possible explanation
for the finite-temperature resistive transition observed recently in
arrays of superconducting grains \cite{baek}.

In conclusion, we have investigated the resistivity scaling of
Josephson-junction arrays at a irrational frustration using a driven
MC dynamics \cite{eg04}. The results are consistent with a
phase-coherence transition scenario where the critical temperature
vanishes, $T_c=0$. The linear resistivity is finite at nonzero
temperatures but nonlinear behavior sets in at a crossover current
determined by the thermal critical exponent $\nu$. The results agree
with earlier simulations using the resistively shunted-junction
model for the dynamics of the array \cite{eg96} and more recent MC
simulations taking into account the intrinsic finite-size effect
\cite{park}. Although we have only studied the array at a particular
value of irrational frustration, the golden mean, we believe that
the conclusion of a zero-temperature phase-coherence transition
should be valid for all irrationals but possibly with different
values of the thermal critical exponent $\nu$. The main advantage of
studying the golden mean value is that it is considered the farthest
from the low-order rationals and so intrinsic finite-effects should
be smaller. However, other irrational frustrations have also been
studied numerically \cite{park,kolahchi} and experimentally
\cite{yu}. The resistive behavior probes mainly the phase-coherence
of the system and since we find that phase coherence is only
attained at zero temperature, we can not address directly the
question of the existence of a vortex-order transition at finite
temperatures. In fact, vortex order does not require long-range
phase coherence. Therefore, a vortex-order transition at zero
temperature or at finite temperature is consistent with the present
work. However, in view of the results for the supercooled relaxation
\cite{kimlee} suggesting an analogy to structural glasses such
transition may be expected at finite temperature and in fact is
consistent with  MC simulations indicating a first-order vortex
transition \cite{teitelf,tang,llkim}. Thus, the interesting
possibility arises where the array undergoes two transitions for
decreasing temperature, a finite-resistance vortex-order transition
at finite temperature and a superconducting transition only at zero
temperature. This phase transition scenario and the predicted
behavior of the linear and nonlinear resistivity provides an
interesting experimental signature for a Josephson-junction array at
irrational frustration.

\medskip

This work was supported by FAPESP (grant 03/00541-0) and computer
facilities from CENAPAD-SP.

\end{document}